%% file: scaling-listnet.tex
\documentclass[sigconf,natbib,9pt]{acmart}

\input{packages}
\input{copyright}
\input{definitions}

\usepackage{setspace}

\title{Modeling Label Ambiguity for \\Neural List-Wise Learning to Rank}

\author{Rolf Jagerman}
\affiliation{%
\institution{University of Amsterdam}
\city{Amsterdam}
\country{The Netherlands}
}
\email{rolf.jagerman@uva.nl}

\author{Julia Kiseleva}
\affiliation{%
	\institution{\hspace{-3.2mm}UserSat.com \& University of Amsterdam}
	\city{Amsterdam}
	\country{The Netherlands}
}
\email{j.kiseleva@uva.nl}

\author{Maarten de Rijke}
\orcid{0000-0002-1086-0202}
\affiliation{
\institution{University of Amsterdam}
\city{Amsterdam}
\country{The Netherlands}
}
\email{derijke@uva.nl}

\begin{document}
	\input{sections/abstract}

\ccsdesc[500]{Computing methodologies~Neural networks}
\ccsdesc[500]{Information systems~Learning to rank}

\keywords{Neural Information Retrieval; List-wise Learning to Rank}

	\maketitle
	
	\input{sections/introduction}
	\input{sections/relatedwork}
	\input{sections/method}

\input{sections/experiments}

	\input{sections/conclusion}


\medskip
\begin{spacing}{1}
\noindent\small
\textbf{Acknowledgments.}
This research was supported by
Ahold Delhaize,
Amsterdam Data Science,
the Bloomberg Research Grant program,
the Criteo Faculty Research Award program,
the Dutch national program COMMIT,
Elsevier,
the European Community's Seventh Framework Programme (FP7/2007-2013) under
grant agreement nr 312827 (VOX-Pol),
the Microsoft Research Ph.D.\ program,
the Netherlands Institute for Sound and Vision,
the Netherlands Organisation for Scientific Research (NWO)
under pro\-ject nrs
612.001.116, 
HOR-11-10, 
CI-14-25, 
652.\-002.\-001, 
612.\-001.\-551, 
652.\-001.\-003, 
and
Yandex.
All content represents the opinion of the authors, which is not necessarily shared or endorsed by their respective employers and/or sponsors.
\end{spacing}

	\bibliographystyle{ACM-Reference-Format}
	\bibliography{scaling-listnet} 
\end{document}

%% file: packages.tex
\PassOptionsToPackage{dvipsnames}{xcolor}

\usepackage{algorithm}
\usepackage{algpseudocode}
\usepackage{graphicx}
\usepackage{amsmath}
\usepackage{multirow}
\usepackage{hyperref}
\usepackage{cleveref}
\usepackage{xcolor}
\usepackage{natbib}
\usepackage{paralist}
\usepackage[T1]{fontenc}
\usepackage{beramono}
\usepackage{listings}
\usepackage[dvipsnames]{xcolor}
\usepackage[inline]{enumitem}
\usepackage{acronym}
\usepackage{bm}
\usepackage{subcaption}
\usepackage{bbm}
\usepackage{xifthen}

%% file: copyright.tex
\copyrightyear{2017}
\acmYear{2017}
\setcopyright{rightsretained}
\acmConference{SIGIR 2017 Workshop on Neural Information Retrieval (Neu-IR'17)}{}{August 7--11, 2017, Shinjuku, Tokyo, Japan} \acmPrice{}

%% file: definitions.tex
\acrodef{IR}{Information Retrieval}
\acrodef{LTR}{Learning to Rank}
\acrodef{PL}{Plackett-Luce}
\acrodef{SGD}{Stochastic Gradient Descent}

\newcommand{\OurMethod}{ListPL}
\newcommand{\RQ}{Can we learn a list-wise neural model while taking into account the ambiguity of relevance labels in \ac{LTR} data?}
\newcommand{\OurSource}{\url{https://github.com/rjagerman/shoelace}}

\newcommand{\docset}{\mathbb{D}}
\newcommand{\doc}{\bm{d}}
\newcommand{\queryset}{\mathbb{Q}}
\newcommand{\query}{q}

\newcommand{\labelset}{\mathbb{Y}}
\newcommand{\labels}{y}
\newcommand{\permset}{\Omega}
\newcommand{\perm}{\pi}
\newcommand{\predictor}[1]{
	\ifthenelse{\isempty{#1}}{
		f
	}{
		f(#1)
	}
}
\newcommand{\loss}[1]{\mathcal{L}\left(#1\right)}
\newcommand{\indicator}[1]{\bm{1}_{#1}}
\newcommand{\deriv}[1]{\frac{\partial}{\partial #1}}
\newcommand{\pl}[2]{PL\left(#1 \mid #2\right)}

\newcommand\doSingleLine[1]
{\let\normalnewline=\\
	\let\\=\relax
	#1%
	\let\\=\normalnewline}

%% file: sections/abstract.tex

\begin{abstract}
	List-wise learning to rank methods are considered to be the state-of-the-art. One of the major problems with these methods is that the ambiguous nature of relevance labels in learning to rank data is ignored. Ambiguity of relevance labels refers to the phenomenon that multiple documents may be assigned the same relevance label for a given query, so that no preference order should be learned for those documents. In this paper we propose a novel sampling technique for computing a list-wise loss that can take into account this ambiguity. We show the effectiveness of the proposed method by training a 3-layer deep neural network. We compare our new loss function to two strong baselines: ListNet and ListMLE. We show that our method generalizes better and significantly outperforms other methods on the validation and test sets.
\end{abstract}

%% file: sections/introduction.tex

\section{Introduction}
One of the most important components to any search engine is the \ac{LTR} model. It considers dozens or even hundreds of relevance signals and determines in what order to show the documents to the user based on these signals. The following three main directions have emerged in the field of \ac{LTR}:
\begin{enumerate}
	\item \textbf{Point-wise}~\cite{fuhr1989optimum,cooper1992probabilistic,crammer2001pranking}: Models the \ac{LTR} problem as a (usually probabilistic) regression problem.
	\item \textbf{Pair-wise}~\cite{joachims2002optimizing,burges2005learning}: Casts the \ac{LTR} problem as a classification problem and learns preferences between pairs of documents.
	\item \textbf{List-wise}~\cite{burges2006learning,cao2007learning,xia2008listwise}: Attempts to solve \ac{LTR} by treating lists of preferences as instances for learning; these methods are considered to be the current state-of-the-art.
\end{enumerate}
In this paper, we focus on list-wise \ac{LTR}, since these methods are the current state-of-the-art and are commonly used in conjunction with neural networks. In particular, we focus on ListNet~\cite{cao2007learning} and ListMLE~\cite{xia2008listwise}. One of the major difficulties with these list-wise methods is that there is no consideration for the ambiguity that exists in \ac{LTR} data that uses relevance scores.

\begin{figure}[h]
	\includegraphics[width=0.9\linewidth]{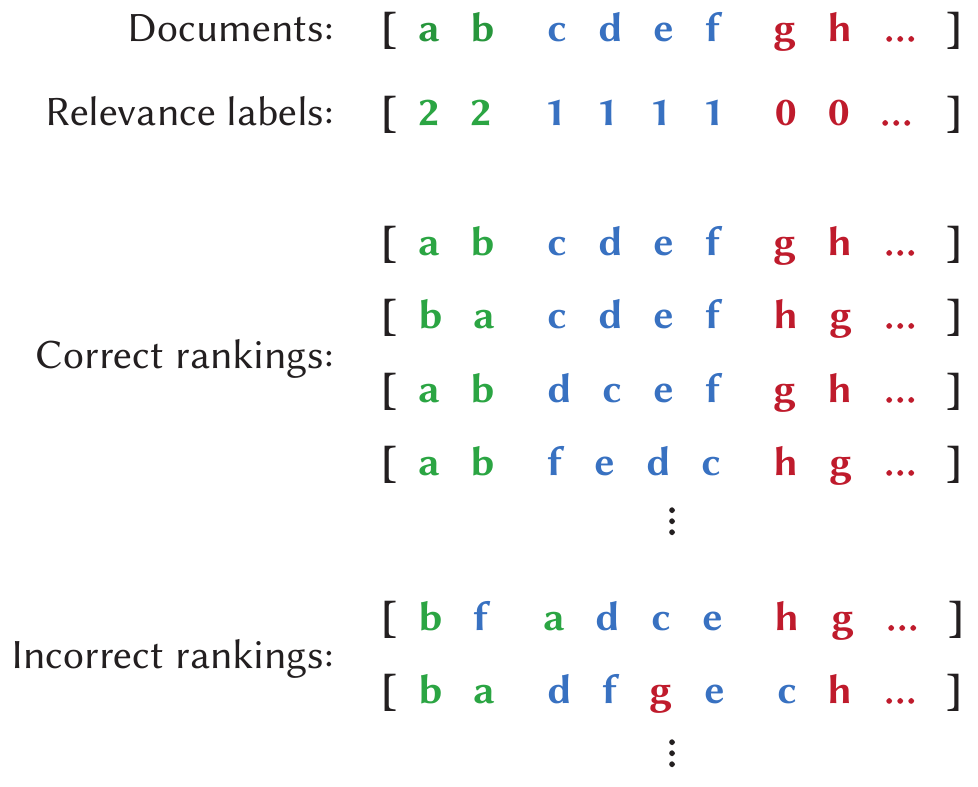}
	\caption{Relevance labels in \ac{LTR} admit many different ``correct'' rankings for the same query $q$. The colors indicate different relevance grades: green is highly relevant, blue is relevant and red is not relevant.}
	\label{fig:ambiguity}
\end{figure}

In \ac{LTR} we deal with relevance labels that score documents on a finite ordinal scale. Let us consider an example in Figure~\ref{fig:ambiguity}.
\\There are typically more documents than relevance labels, which necessarily introduces ambiguity in the rankings. Any documents that share the same relevance label can freely be interchanged, thus any permutation of documents with the same relevance label is technically correct. As a consequence, there are many possible rankings of the documents that would be considered ``correct'' for the same 	query $q$. That brings us to the problem of \emph{label ambiguity} which is the phenomenon that multiple documents may be assigned the same relevance label for a given query, so that no preference order should be learned for those documents. Learning a preference where none exists may lead to overfitting or limitations in the learner's ability to generalize. 

In ListNet, it is computationally too expensive to model label ambiguity, because every possible permutation has to be considered. \citet{cao2007learning} address this problem by introducing a top-$k$ approximation. However, we argue that this largely mitigates the major attractiveness of ListNet, namely its capability to learn from the full ranked list. For ListMLE, \citet{xia2008listwise} make a simplifying assumption by sampling one perfect ranking and assuming that this is the ground truth. These drawbacks lead to our main research question:

\noindent
\begin{center}
\vspace{0.1cm}
\mbox{
	\parbox{0.83\columnwidth}{
		\textit{\RQ}
	}
}
\vspace{0.1cm}
\end{center}

\noindent%
Our contributions are two-fold:
\begin{itemize}
	\item We introduce a novel way to optimize a list-wise neural \ac{LTR} model, sampling learning instances directly from the Plackett-Luce distribution~\cite{plackett1975analysis,luce2005individual}, to take into account the ambiguity of relevance labels in an \ac{LTR} setting.
	\item We publish the source code of our method, which uses Chainer~\cite{chainer_learningsys2015}, a GPU-accelerated deep learning framework, promoting future research in neural list-wise \ac{LTR} methods.\footnote{\OurSource}
\end{itemize}

\noindent%
The remainder of this paper is organized as follows. 
Section~\ref{sec:rel_work} introduces our notation and  earlier work on list-wise \ac{LTR} approaches.
Section~\ref{sec:label_ambigiuty} discusses in details the phenomena of label ambiguity. We present our solution in Section~\ref{sec:method}. 
We present our experimental results in Section~\ref{sec:experiments}.
Finally, we conclude in Section~\ref{sec:conclusion}.

%% file: sections/relatedwork.tex

\section{Related work}
\label{sec:rel_work}

In this section we discuss several important works that our method builds on. First, we introduce some notation that is used throughout this paper (Section~\ref{sec:preliminaries}). Next, we briefly discuss the \ac{PL} distribution (Section~\ref{sec:pld}), followed by ListNet~\cite{cao2007learning} (Section~\ref{sec:listnet}) and ListMLE~\cite{xia2008listwise} (Section~\ref{sec:listmle}).

\subsection{Preliminaries}
\label{sec:preliminaries}
We consider the scenario of \acf{LTR} where we are given a collection of $m$ queries $\queryset = \{\query^{(1)}, \ldots, \query^{(m)}\}$. Each query $\query^{(i)}$ is associated with a set of $n$ documents $\docset^{(i)} = \{\doc^{(i)}_1, \ldots, \doc^{(i)}_n\}$ and corresponding relevance labels $\labelset^{(i)} = \{\labels^{(i)}_1, \ldots, \labels^{(i)}_n\}$. Each document $\doc^{(i)}_j$ is a $d$-dimensional feature vector representing the query-document pair. For the sake of brevity, we will drop the superscript notation $\cdot^{(i)}$ for the remainder of this paper.

In \ac{LTR} we use a scoring function $\predictor{}$ to score every document $\doc_j$ and then sort the set of documents by these scores. Our objective is to find a function $\predictor{}$ for which the resulting ranking is optimal with regards to the relevance labels. In other words, we wish to find a function $\predictor{}$ that assigns high scores to documents that have high relevance and low scores to documents that have little relevance.

\subsection{Plackett-Luce Distribution}
\label{sec:pld}
The \acf{PL} distribution~\cite{plackett1975analysis,luce2005individual} is used extensively in probabilistic list-wise \ac{LTR} methods. It is based on the idea that a ranking is drawn sequentially from a list of item-specific scores, one item at a time. The \ac{PL} distribution is a probability distribution over all possible permutations of a set of item scores. Intuitively, it assigns a high probability to permutations that place high-scoring items at the top while assigning low probability to permutations that place high-scoring items at the bottom. More formally, the \ac{PL} probability for a permutation $\perm = \{\perm_1, \ldots, \perm_n\}$, given a scoring function $\predictor{}$ and a set of documents $\docset = \{\doc_1, \ldots, \doc_n\}$ is defined as follows:
\begin{equation}
\pl{\perm}{\docset; \predictor{}} = \prod_{i=1}^{n} \frac{\phi(\predictor{\doc_{\perm_i}})}{\sum_{j=i}^{n} \phi(\predictor{\doc_{\perm_j})})},
\end{equation}
where $\phi : \mathbb{R} \rightarrow \mathbb{R}_{>0}$ is a function that maps real-valued scores to positive numbers. In practice, this is usually chosen to be the exponential function. For the sake of simplifying the derivations in Section~\ref{sec:method}, we also choose the exponential function. Hence, we will from now on assume $\phi(\cdot) = \exp(\cdot)$.

We consider two \ac{LTR} methods that build on the \ac{PL} distribution: ListNet~\cite{cao2007learning} and ListMLE~\cite{xia2008listwise}. It can be shown that the losses associated with these methods are continuous, convex and differentiable~\cite{xia2008listwise}. This makes them easy to optimize via \ac{SGD} and attractive in a neural setting.

\subsection{ListNet}
\label{sec:listnet}
ListNet~\cite{cao2007learning} is one of the earliest list-wise \ac{LTR} methods. It considers the following two \ac{PL} distributions:
\begin{itemize}
	\item $\pl{\perm}{\docset; \predictor{}}$\\This is the \ac{PL} distribution of the output of the network $\predictor{}$. Any permutation $\perm$ that places documents that generated large network scores at the top gets assigned high probability.
	\item $\pl{\perm}{\docset; \psi_{\labelset}}$\\This is the \ac{PL} distribution of a mapping of the relevance labels $\labelset$. The mapping $\psi_{\labelset}$ generates a score vector of the ground truth that retains the order of the relevance labels:
	\begin{equation}
	\psi_{\labelset}(\labels_i) > \psi_{\labelset}(\labels_j) \iff \labels_i > \labels_j
	\end{equation} Any permutation $\perm$ that places documents with high relevance at the top gets assigned high probability.
\end{itemize}
The optimization objective then becomes minimizing the Kullback-Leibler divergence of these two distributions:
\begin{equation}
\min_f D_{\text{KL}}\left(\pl{\perm}{\docset; \psi_{\labelset}}~||~\pl{\perm}{\docset; \predictor{}} \right)
\end{equation}

This is equivalent to minimizing the cross entropy, giving rise to the following loss function:
\begin{equation}
\loss{\predictor{\docset}, \labelset} = - \sum_{\perm \in \permset} \pl{\perm}{\docset; \psi_{\labelset}} \log \left(\pl{\perm}{\docset; \predictor{}}  \right),
\label{eq:listnet}
\end{equation}
where $\permset$ is the set of all possible permutations.

To compute this cross entropy loss, we have to consider every possible permutation in $\permset$, which is of the size $\mathcal{O}(n!)$. \citet{cao2007learning} resort to a top-$k$ approximation, where $k$ is usually 1, to make the computation feasible.

A stochastic top-$k$ ListNet variant has been proposed by~\citet{luo2015}. In this paper, the authors sample from within the top-$k$ subgroups to speed up training. This is different to our work, where we completely eliminate the need for a top-$k$ approximation.

\subsection{ListMLE}
\label{sec:listmle}
ListMLE~\cite{xia2008listwise} replaces the optimization objective of ListNet with a simpler form. The simplifying assumption is that a single permutation $\pi \in \{\perm \mid y_{\pi_i} \geq y_{\pi_j}; i < j\}$ is chosen and considered to be \emph{the ground truth}. It then directly optimizes the \ac{PL} probability of the network scores for this permutation and uses the negative log probability as a loss:
\begin{equation}
\loss{\predictor{\docset}, \labelset} = - \log \pl{\perm}{\docset; \predictor{}}.
\label{eq:listmle}
\end{equation}
One of the main drawbacks of ListMLE is that it assumes that a single perfect ranking $\perm$ is known. This assumption however does not hold for \ac{LTR} data sets, where we have ambiguous relevance labels.

\medskip\noindent%
To summarize, both ListNet and ListMLE build on the \ac{PL} distribution and provide elegant probabilistic ways to learn a list-wise \ac{LTR} model. 
The key distinction of our work compared to previous efforts is that we introduce a new \ac{LTR} method that properly deals with the \emph{label ambiguity} problem, which we will describe next.

%% file: sections/method.tex

\section{The Ambiguity of Ranking}
\label{sec:label_ambigiuty}

Existing work on list-wise approaches typically ignores the ambiguity of the labels. For instance, ListMLE samples a \emph{single} permutation from the ground truth, whose ordering is then assumed to be \emph{the} ground truth ranked list. For example, take the following $8$ documents with corresponding relevance scores:
\begin{equation}
\begin{array}{l@{}l@{}l}
	\begin{array}{r}
		\docset = [ \\
		\labelset = [
	\end{array}
	&
	\begin{array}{@{}cccccccc@{}}
	\doc_1 & \doc_2 & \doc_3 & \doc_4 & \doc_5 & \doc_6 & \doc_7 & \doc_8 \\
	2 & 2 & 1 & 1 & 1 & 0 & 0 & 0
	\end{array}
	&
	\begin{array}{l}
		] \\
		]
	\end{array}
\end{array}
\end{equation}
What we wish to learn are the following preferences:
\begin{equation}
\left\{\doc_1 \leftrightarrow \doc_2\right\} \succ \left\{\doc_3 \leftrightarrow \doc_4 \leftrightarrow \doc_5\right\} \succ \left\{\doc_6 \leftrightarrow \doc_7 \leftrightarrow \doc_8\right\}.
\end{equation}
Instead, the optimization objective in ListMLE learns the following:
\begin{equation}
\doc_1 \succ \doc_2 \succ \doc_3 \succ \doc_4 \succ \doc_5 \succ \doc_6 \succ \doc_7 \succ \doc_8.
\end{equation}
Thus, the learning algorithm will attempt to learn $\doc_1 \succ \doc_2$. This is problematic, because according to the ground truth, the relative ordering of $\doc_1$ and $\doc_2$ is not meaningful. Attempting to learn such overly specific relations is harmful to the generalization power of the learning algorithm.

\medskip\noindent%
Next, we will present our method that overcomes the described issue.



\section{Sampling Rankings from the Plackett-Luce Distribution}
\label{sec:method}

Instead of naively choosing a single permutation of the documents $\perm$ and considering that permutation to be the ground truth, we propose a more sophisticated sampling method. The main idea is to directly sample a ranking from the \ac{PL} distribution of the relevance labels during every stochastic update.

To motivate this decision from a theoretical point of view, let us revisit the ListNet cross entropy loss function. We can rewrite Equation~\ref{eq:listnet} into the following form:
\begin{equation}
\loss{\predictor{\docset}, \labelset} = \sum_{\perm \in \permset} \underbrace{\pl{\perm}{\docset; \psi_{\labelset}}}_{\text{weight}} \underbrace{ \left( - \log \left(\pl{\perm}{\docset; \predictor{}}  \right) \right) }_{\text{loss}}
\end{equation}
This loss can be interpreted as a stochastic variant of the ListMLE loss. Here we sample a possible permutation $\perm$ with a corresponding probability $\pl{\perm}{\docset; \psi_{\labelset}}$. We can then use that sample to compute a stochastic loss. We call this method \OurMethod:
\begin{align}
\begin{aligned}
\loss{\predictor{\docset}, \labelset} = &~- \log \left(\pl{\perm}{\docset; \predictor{}}  \right)\\
\perm \sim &~\pl{\perm}{\docset; \psi_{\labelset}}
\end{aligned}
\label{eq:listpl}
\end{align}
Deriving the stochastic gradient of this loss function follows the same derivation as ListMLE. We include the derivative with respect to the activation function here for completeness sake:
\begin{align}
& \deriv{\predictor{\doc_{\perm_k}}} \left[ - \log \left( \pl{\perm}{\docset; \predictor{}} \right) \right] \nonumber \\
&= - \deriv{\predictor{\doc_{\perm_k}}} \left[ \log \prod_{i=1}^{n} \frac{\exp(\predictor{\doc_{\perm_i}})}{\sum_{j=i}^n \exp(\predictor{\doc_{\perm_j}})} \right] \nonumber \\
&= - \deriv{\predictor{\doc_{\perm_k}}} \left[ \sum_{i=1}^n \left( \predictor{\doc_{\perm_i}} - \log \sum_{j=i}^n \exp(\predictor{\doc_{\perm_j}}) \right)\right] \nonumber \\
&= \sum_{i=1}^n \indicator{i \leq k \leq n} \left( \frac{\exp(\predictor{\doc_{\perm_k}})}{\sum_{j=i}^n \exp(\predictor{\doc_{\perm_j}})} \right) - 1 \nonumber \\
&= \sum_{i=k}^n \left( \frac{\exp(\predictor{\doc_{\perm_k}})}{\sum_{j=i}^n \exp(\predictor{\doc_{\perm_j}})} \right) - 1
\end{align}
The resulting loss and corresponding gradient can then be used to train a neural network via \ac{SGD}.

A different way to look at \OurMethod{} is like a data set generation method. We are essentially applying ListMLE to a data set that contains all permutations $\perm$ of possible rankings and weighing each one by $\pl{\perm}{\docset; \psi_{\labelset}}$. Looking at the method from this perspective, it is reasonable to assume we can do sampling-based \ac{SGD} the usual way: sample a ranking $\pi$ uniformly and then weigh the gradient by $\pl{\perm}{\docset; \psi_{\labelset}}$. However, this runs into a problem: the sample space is enormous ($\mathcal{O}(n!)$). With sufficiently many documents per query, most uniformly sampled rankings will have a \ac{PL} probability that is close to 0 resulting in a very slow convergence rate.

\medskip\noindent%
In this section, we described our method, \OurMethod{}, which is able to deal with the label ambiguity problem described in Section~\ref{sec:label_ambigiuty}. Next, we will present our experimental evaluation.

%% file: sections/experiments.tex
\section{Experiments and Results}
\label{sec:experiments}

\begin{figure*}
	\centering
	\vspace{-0.1cm}
	\begin{subfigure}[t]{0.325\textwidth}
		\centering
		\includegraphics[clip,trim=2mm 2mm 3mm 0mm,width=\textwidth]{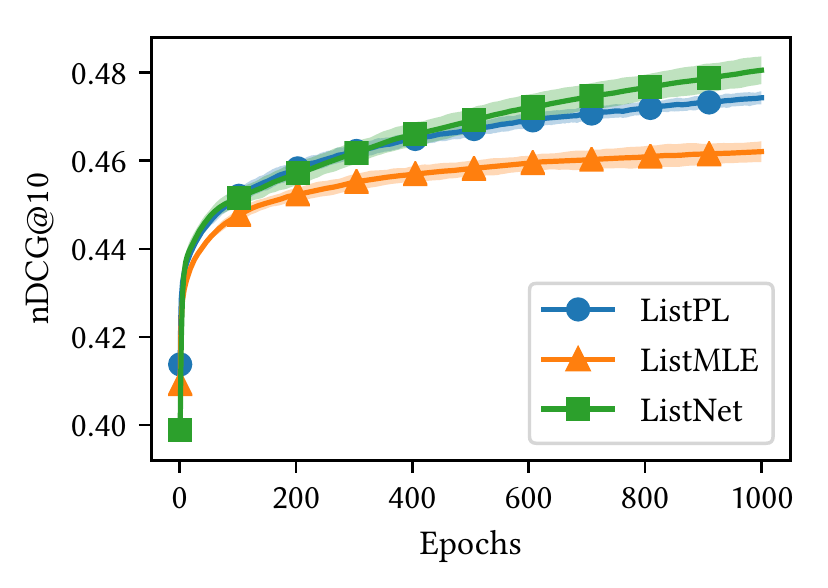}
		\caption{Training set}
		\label{fig:ndcg-results-training}		
	\end{subfigure}
	~ 
	\begin{subfigure}[t]{0.325\textwidth}
		\centering
		\includegraphics[clip,trim=2mm 2mm 3mm 0mm,width=\textwidth]{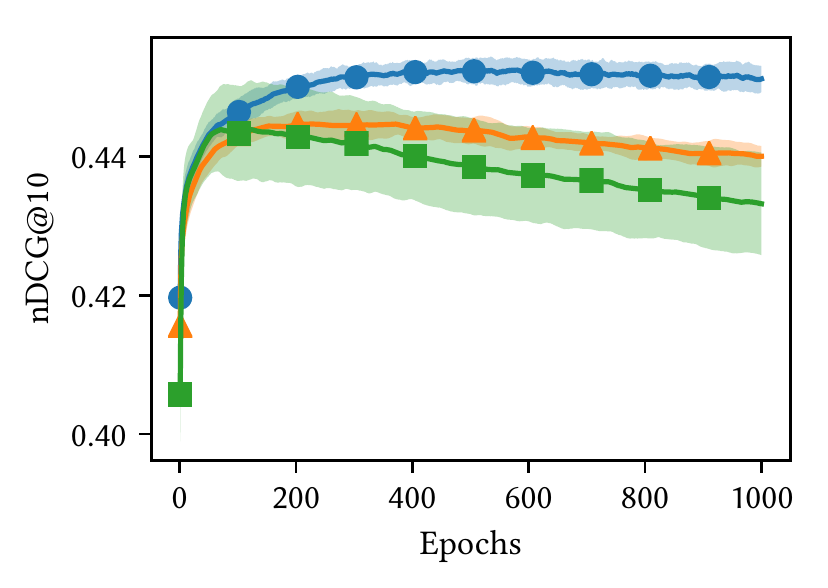}
		\caption{Validation set}
		\label{fig:ndcg-results-validation}		
	\end{subfigure}
	~
	\begin{subfigure}[t]{0.325\textwidth}
		\centering
		\includegraphics[clip,trim=2mm 2mm 3mm 0mm,width=\textwidth]{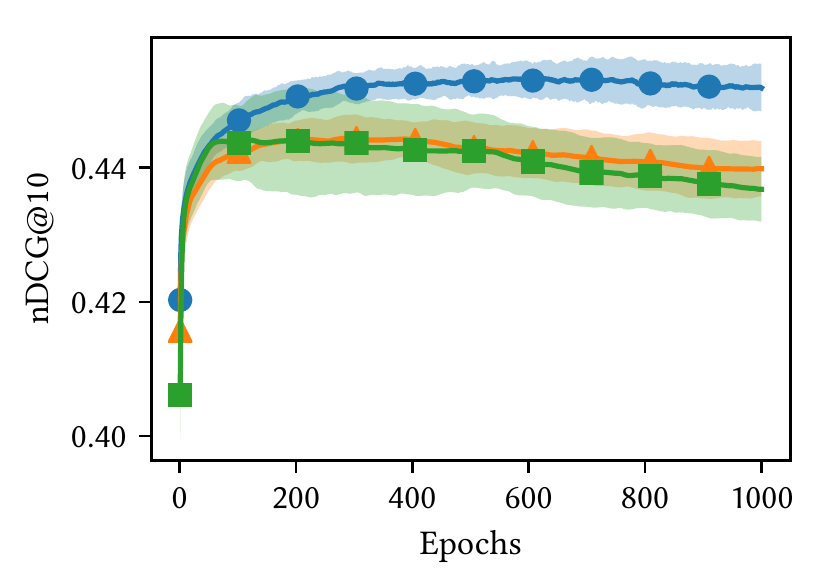}
		\caption{Test set}
		\label{fig:ndcg-results-test}		
	\end{subfigure}
	\caption{nDCG@10 performance on MSLR-WEB10K. The shaded areas indicate the standard deviation over the different folds. On the test set, the performance improvements over ListNet and ListMLE are statistically significant with $p = 0.00078$ and $p = 0.00218$ respectively (two-tailed t-test).}
	\label{fig:ndcg-results}
\end{figure*}

To validate the effectiveness of our method, \OurMethod, we compare it to ListNet (top-1) and ListMLE on the MSLR-WEB10k data set~\cite{DBLP:journals/corr/QinL13}. This data set contains 10,000 queries. It represents query-document pairs as 136-dimensional feature vectors and grades them on a scale from 0 (irrelevant) to 4 (perfectly relevant). We use 6,000 queries for training, 2,000 for validation and 2,000 for testing.

The architecture of our network is a 3-layer fully connected neural network with 80 ReLU activation units at each hidden layer. We experimented with more than 3 layers and found only negligible improvements. We keep the network architecture the same and vary only the loss function:
\begin{itemize}
	\item ListNet loss using a top-1 approximation (Equation \ref{eq:listnet})
	\item ListMLE loss (Equation \ref{eq:listmle})
	\item ListPL loss (Equation \ref{eq:listpl})
\end{itemize}
ADAM~\cite{kingma2014adam} is used as the optimizer with the default parameters $\beta_1 = 0.9$, $\beta_2 = 0.999$ and a learning rate $\alpha = 10^{-5}$. The experiments are run for 1000 epochs.

Figure~\ref{fig:ndcg-results} shows the results on the MSLR-WEB10k data set. We evaluate the performance of the methods using nDCG@10~\cite{jarvelin2000ir}, which is a natural evaluation metric for \ac{LTR} in the Web-search setting. We see that \OurMethod~performs similar to ListNet during training, but outperforms both ListNet and ListMLE during validation and testing. The performance degradations on the validation set and test set for ListNet and ListMLE that occur after 100 epochs indicate that these methods are overfitting and are effectively learning noise coming from label ambiguity. These results are in line with our expectations because \OurMethod{} properly deals with the ambiguity in the relevance scores and thus generalizes better.

The nDCG@10 performance on the test set is evaluated using 5-fold cross validation. The performance improvement of \OurMethod~over ListNet is statistically significant with $p = 0.00078$ (two-tailed t-test) whereas the performance improvement over ListMLE is statistically significant with $p = 0.00218$ (two-tailed t-test).

\medskip\noindent%
To summarize, based on extensive experimentation with the MSLR-WEB10k data set, we conclude that \OurMethod{} significantly outperforms strong baselines due to the fact that it handles the label ambiguity problem well, and, hence, generalizes better.

%% file: sections/conclusion.tex

\section{Conclusion}
\label{sec:conclusion}

The paper extends earlier work on list-wise approaches for \ac{LTR}~\citep{burges2006learning,cao2007learning,xia2008listwise}. Specifically,  we have considered the problem of label ambiguity. 
Our main research question was:
\noindent
\begin{center}
	\vspace{0.1cm}
	\mbox{
		\parbox{0.83\columnwidth}{
			\textit{\RQ}
		}
	}
	\vspace{0.1cm}
\end{center}
Our overall conclusion is that by introducing a sampling method based on directly sampling from the \acf{PL} distribution of relevance labels, we are able to increase the ability to generalize neural list-wise \ac{LTR} methods while maintaining efficiency. 
Specifically, we used a modified loss function that can efficiently mitigate the problem of label ambiguity and thereby improve over existing list-wise neural \ac{LTR} methods. 

Our extensive experimentation with the MSLR-WEB10k data set showed that our method, \OurMethod{}, significantly outperforms  two strong baselines: ListNet~\cite{cao2007learning} and ListMLE~\cite{xia2008listwise}.
Our method and baselines are implemented using Chainer~\cite{chainer_learningsys2015}, a GPU-accelerated deep learning framework. We are sharing the source code of \OurMethod{} online\footnote{\OurSource} (MIT licensed), which we hope is useful for future research towards list-wise neural methods.